\documentclass{mycls}

\usepackage{natbib}

\usepackage[dvips]{graphicx}
\usepackage{psfrag}
\usepackage{epsfig}
\usepackage{amsmath,rotating}

\graphicspath{{figures/}}
\usepackage{color}

\title{Jet-edge interaction tones}

\shorttitle{Jet-edge interaction tones}

\title{Jet-edge interaction tones}
\author{ \text{P. Jordan$^1$},
  V. Jaunet$^1$,
 A. Towne$^2$,
 A.V.G. Cavalieri$^3$,\\
 T. Colonius$^4$,
O. Schmidt$^4$,
 A. Agarwal$^5$\\
 \vspace{0.5cm}
 {}$^1$ {\normalfont \small{D\' epartement Fluides, Thermique, Combustion, Institut PPRIME,   \\ CNRS---Universit\'e de Poitiers---ENSMA, Poitiers, France}} \\
 \vspace{0.1cm}
 {}$^2$ {\normalfont \small{Center for Turbulence Research, Stanford University, Stanford, CA 94305, USA}} \\
 \vspace{0.1cm}
 {}$^3$ {\normalfont \small{Divis{\~a}o de Engenharia Aeron{\'a}utica, Instituto Tecnol{\'o}gico de Aeron{\'a}utica, \\ S{\~a}o Jos{\'e} dos Campos, SP, Brazil}}\\
 \vspace{0.1cm}
 {}$^4$ {\normalfont \small{Division of Engineering and Applied Science, California Institute of Technology, \\ Pasadena, CA 91125, USA}}\\
 \vspace{0.1cm}
 {}$^5$ {\normalfont \small{Department of Engineering, University of Cambridge, UK}}
}

\shortauthor{Jordan et~al..}

\begin{document}

\maketitle

\noindent {\textbf{Abstract.} }\textcolor{black}{Motivated by the problem of jet-flap interaction noise, we study the tonal dynamics that occur when a sharp edge is placed in the hydrodynamic nearfield of an isothermal turbulent jet. We perform hydrodynamic and acoustic pressure measurements in order to characterise the tones as a function of Mach number and streamwise edge position. The distribution of spectral peaks observed, as a function of Mach number, cannot be explained using the usual edge-tone scenario, in which resonance is underpinned by coupling between downstream-travelling Kelvin-Helmholtz wavepackets and upstream-travelling sound waves. We show, rather, that the strongest tones are due to coupling between the former and upstream-travelling jet modes recently studied by \citet{TowneetalJFM2017} and \citet{SchmidtetalJFM2017}. } \textcolor{black}{We also study the band-limited nature of the resonance, showing a high-frequency cut-off to be due to the frequency dependence of the upstream-travelling waves. At high Mach number these become evanescent above a certain frequency, whereas at low Mach number they become progressively trapped with increasing frequency, a consequence of which is their not being reflected in the nozzle plane.} \textcolor{black}{Additionally, a weaker, low-frequency, forced-resonance regime is identified that involves the same upstream travelling  jet modes but that  couple, in this instance, with downstream-travelling sound waves. It is suggested that the existence of two resonance regimes may be due to the non-modal nature of wavepacket dynamics at low-frequency.} 

\section{Introduction}

\textcolor{black}{Resonant phenomena are widely encountered in fluid systems, where they are underpinned by diverse flow physics. They may be exploited to produce pleasant, desired effects, as is the case with musical wind-instruments \citep{howe1975contributions,coltman1976jet,fabre2012aeroacoustics}. Or they may constitute an undesired behaviour that complicates the design of engineering systems. This can occur for flow in the presence of sharp edges \citep{richardson1931edge,powell1953edge,curle1953mechanics}  or cavities \citep{rossiter1964wind,rowley2002self,kegerise2004mode}; it is the case for imperfectly expanded supersonic jets, that `screech' \citep{powell1953mechanism,alkislar2003structure,edgington2014coherent}, impinging jets \citep{powell1988sound,krothapalli1999flow,henderson2005experimental} and globally unstable flows more generally \citep{huerre1990local,monkewitz1993global}.}

\textcolor{black}{Some kind of feedback is usually at work when a fluid system undergoes resonance. Typically this involves a disturbance, initiated at some point in the flow, and that travels in a given direction, triggering, at a distant point, a second disturbance that travels back toward the inception point of the first. Synchronisation of the two can occur if their phases are appropriately matched at the inception and reflection points. In this `long-range' feedback scenario, the inception and reflection points may correspond to physical boundaries, as is the case in cavity flows, or they may arise due to other flow phenomena, such as shocks in underexpanded supersonic jets, or turning points in slowly spreading mean-flows \citep{rienstra2003sound,TowneetalJFM2017}. In certain globally unstable flows, resonance may occur in the absence of solid boundaries, between disturbances of opposite generalised group velocity when their frequencies and wavenumbers become matched; this is the case for the saddle-point ringing that underpins absolute instability in wake flows or low-density jets for instance \citep{huerre1990local}.}

\textcolor{black}{In many of the examples evoked above, the downstream-travelling disturbance is a convectively unstable Kelvin-Helmholtz wave, and the upstream-travelling disturbance a sound wave. But in the case of round jets there are other kinds of wave available for both upstream and downstream transport of fluctuation energy. \citet{tam1989three} discuss one such upstream-travelling wave, originally observed by \citet{michalke1970note}, who disregarded it as an artefact of the locally parallel framework of his analysis. It has been suggested by \citet{tam1990theoretical} that this wave may be important in explaining the tonal behaviour of impinging subsonic jets, and a recent numerical study by \citet{bogey2017feedback} shows that this may also be the case for resonance in impinging supersonic jets. \cite{TowneetalJFM2017} and \citet{SchmidtetalJFM2017} recently discovered that the round jet can support a number of additional waves. Resonance possibilities in jets are therefore more numerous than had previously been thought.}

\textcolor{black}{The present study was motivated by the new generation of Ultra-High-Bypass-Ratio turbofan engines and the potential problems posed by the closely coupled jet-flap configurations that such systems involve. With this in mind, we consider the problem of subsonic jets grazing a sharp edge. Such a configuration was studied by  \cite{lawrence2015installed}, who observed a tonal behaviour that could not be explained in terms of upstream-travelling, free-stream sound waves. We perform} similar experiments, involving a flat rectanguler plate whose edge is positioned in the hydrodynamic nearfield of a round turbulent jet. The nozzle is the same used in recent work \citep{cavalieri2013wavepackets,bres2015large,jordan2017modal,jaunet2017two,TowneetalJFM2017,SchmidtetalJFM2017} and among the exit conditions considered are those of the cited studies. Our objective is to establish if the strong hydrodynamic and acoustic tones that are observed in this closely coupled jet-edge configuration can be understood in the framework of the waves considered by \cite{TowneetalJFM2017} and \citet{SchmidtetalJFM2017}.


\begin{figure}
\begin{center}
\begin{tabular}{cc}
\includegraphics[width=0.45\textwidth,clip=]{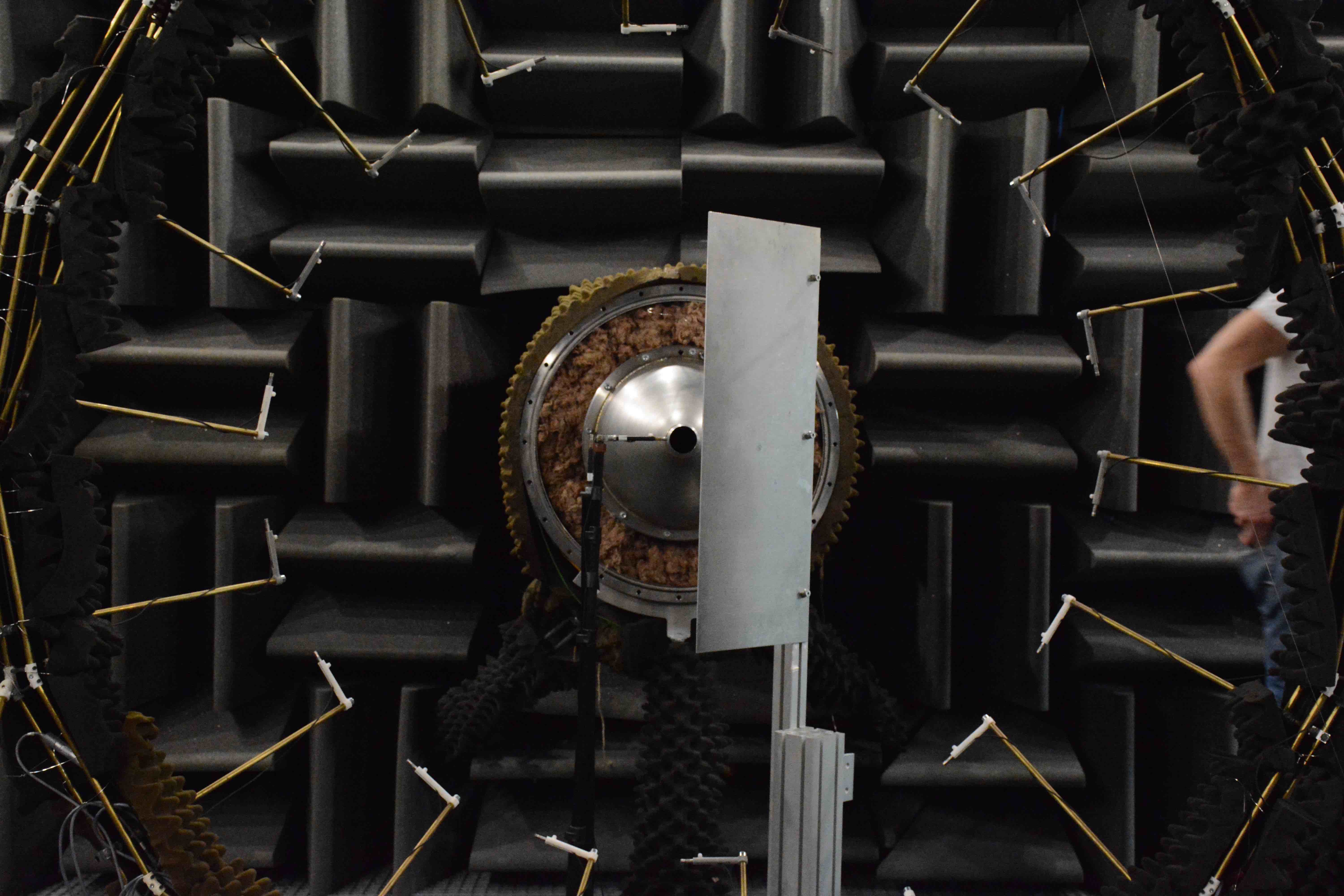} & \includegraphics[width=0.45\textwidth,clip=]{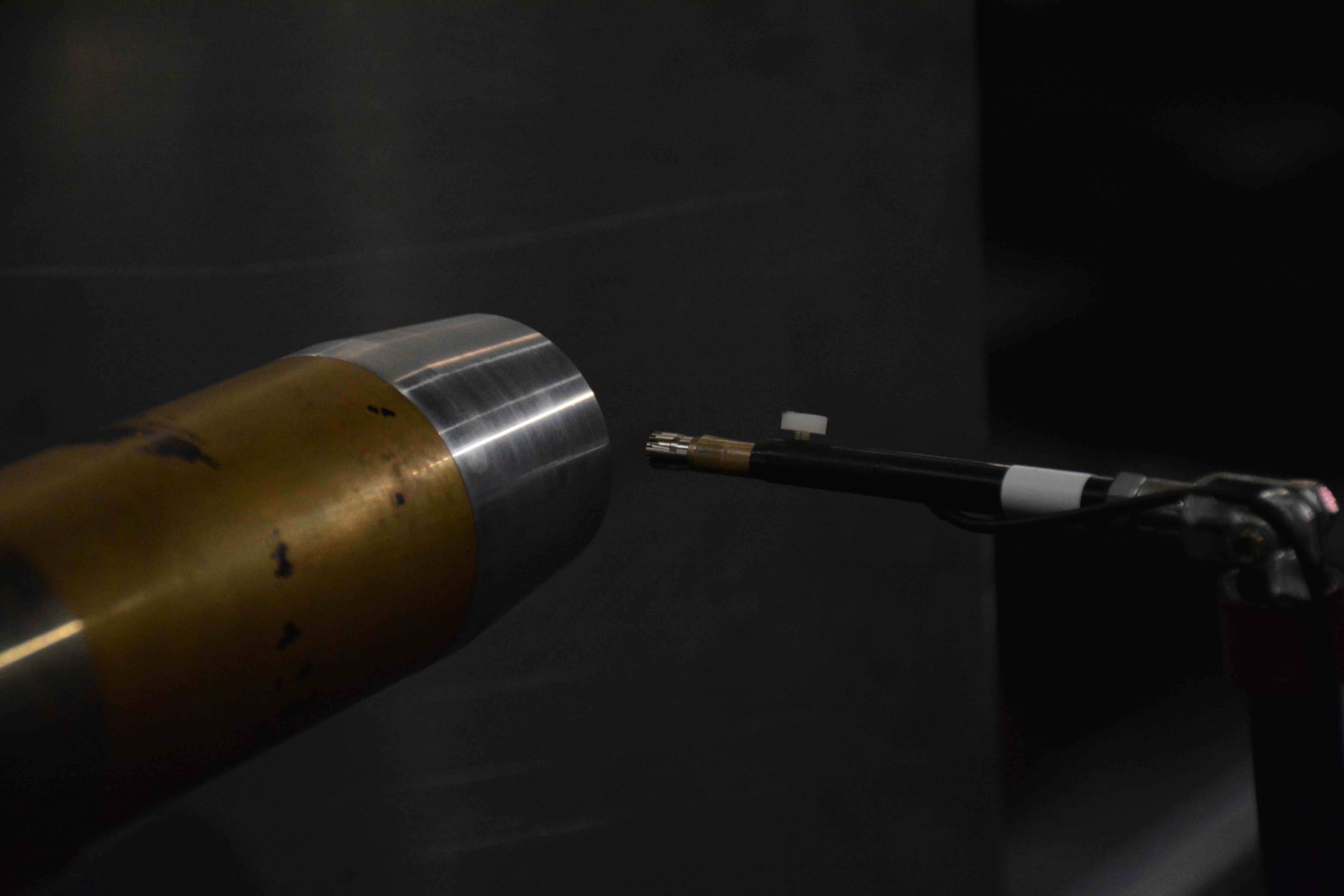} \\
(a) & (b)
\end{tabular}
\caption{Experimental setup: (a) front view showing nozzle, plate, nearfield microphone and azimuthal array of 18 microphones in the acoustic field; (b) close-up view of the nearfield microphone.}\label{fig:expsetup}
\end{center}
\end{figure}

The remainder of the paper is organised as follows. We present the experimental setup in section \S \ref{sec:exp}. This is followed, in section \S \ref{sec:data}, by an overview of the hydrodynamic and acoustic pressure fields that result from interaction of the turbulent jet, whose Mach number is varied, with the plate edge, whose streamwise position is varied. In section \S \ref{sec:theory} we recall briefly the theoretical framework established by \cite{TowneetalJFM2017} and then use this 
to explain the observed tones. Some concluding remarks are provided in section \S \ref{sec:conclusion}.


\section{Experimental setup}\label{sec:exp}

The experiments were performed at the \textit{Bruit et Vent} jet-noise facility of the PPRIME Institute, Poitiers, France. The experimental setup, shown in figure \ref{fig:expsetup}, involved a rectanguler plate situated in the nearfield of a round, isothermal jet of diameter, $D=0.05m$. The boundary layer inside the nozzle is turbulent due to tripping by a carborundum strip situated 2.7$D$ upstream of the exit plane \citep{cavalieri2013wavepackets}.  

\textcolor{black}{In the nearfield, microphone and plate positions are described using a cylindrical coordinate system, $(x,r,\phi)$, where $x$, $r$ and $\phi$ refer, respectively, to streamwise, radial and azimuthal positions. In the acoustic field, a spherical coordinate system, centered on the jet centerline in the nozzle exit plane, is used, $(\theta,\phi,R)$, where $\theta$ is measured from the downstream jet axis, and $\phi$ is measured in the clockwise direction with the jet viewed from the downstream, $\phi=0$ being vertically above the jet. All distances are non-dimensionalised using the jet diameter, $D$.}  

The plate was inclined to form an angle of $45^\circ$ with the jet axis; its edge was situated at $r/D=0.5$ and the streamwise position varied from $x/D=2$ to $x/D=4$ in steps of $1D$. For each plate position, the Mach number of the jet was varied from $M=0.6$ to $M=1$ in increments of $\Delta M=0.02$. Pressure measurements were performed for each configuration using a microphone located at $(x/D,r/D)=(0.08,0.55)$, shown in figure \ref{fig:expsetup}(b), in order to record the hydrodynamic nearfield pressure signature (all references to `nearfield' data correspond to measurements provided by this microphone), and by means of an azimuthal array of 18 microphones situated in the acoustic field, at $R/D=14.6$ and $\theta=80^\circ$, as shown in figure \ref{fig:expsetup}. Twenty seconds of data were recorded at a sample rate of 200kHz and power spectral densities were estimated from this data using Welch's method. \textcolor{black}{Frequency is expressed in non-dimensional form in terms of the Strouhal number, $St=fD/U_j$, where $f$ is frequency in $Hz$, and $U_j$ the jet exit velocity in $ms^{-1}$.}



\section{Acoustic and hydrodynamic tones}\label{sec:data}

\begin{figure}
\begin{center}
{\includegraphics[trim=0 0 0 0, clip, width=1\textwidth]{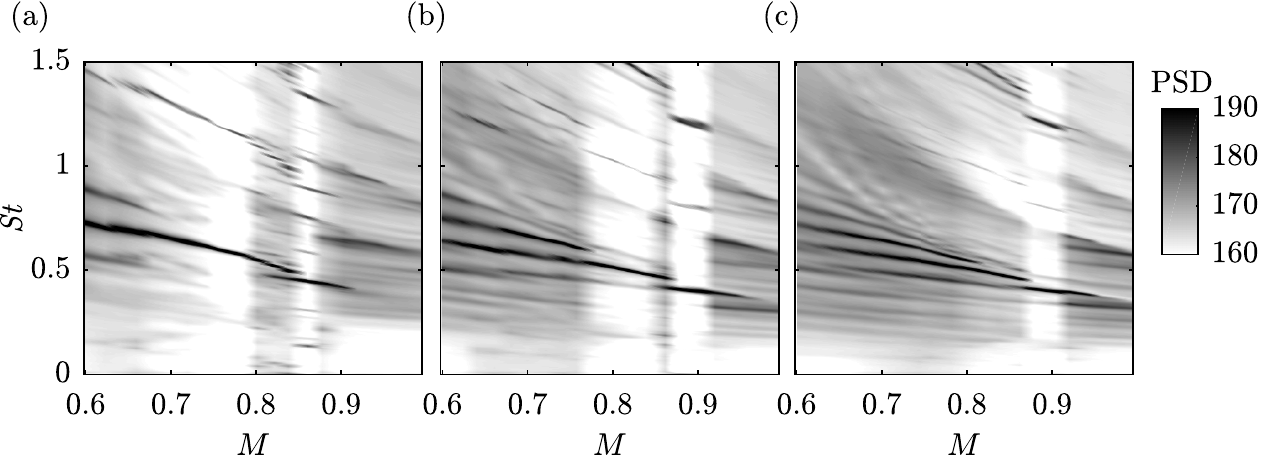}}
\caption{Power-spectral-density maps estimated from hydrodynamic, nearfield pressure recordings for three streamwise edge positions, from left to right: $L/D=2$, $3$ \& $4$.}\label{fig:tones}
\end{center}
\end{figure}

\begin{figure}
\begin{center}
{\includegraphics[trim=0 0 0 0, clip, width=1\textwidth]{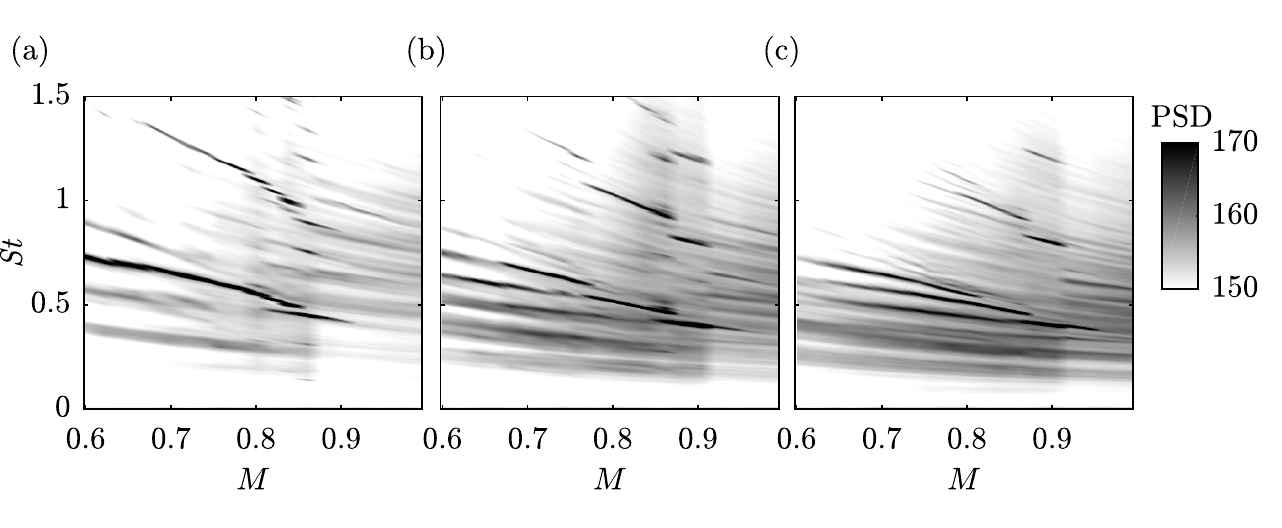}}
\caption{Power-spectral-density maps estimated from acoustic pressure recordings on the `shielded' side of the plate, $(\theta,\phi,R)=(80^\circ,70^\circ,14.6)$, and for three the same streamwise edge positions as figure \ref{fig:tones}.}\label{fig:acou}
\end{center}
\end{figure}

The high sample rate and finely resolved Mach-number variation allow us to obtain high-resolution power-spectral-density (PSD) maps (shown in Figures \ref{fig:tones} and \ref{fig:acou} for the hydrodynamic and acoustic regions, respectively) \textcolor{black}{that comprise a rich ensemble of spectral tones. Peak levels in the acoustic field are of order 170$dB/St$, 
while levels in the hydrodynamic field are, naturally, considerably higher. }

\textcolor{black}{The strongest tones have a similar $St-M$ pattern in both acoustic and hydrodynamic regions, which does not correspond to the usual edge-tone scenario in which resonance occurs between downstream-travelling Kelvin-Helmholtz wavepackets and upstream travelling sound waves.  Following \citet{TowneetalJFM2017}, we use the terms `upstream-' and `downstream-travelling' to refer to the sign of the generalised group velocity \citep{briggs1964electron} of a wave, rather than its phase velocity.} 


\textcolor{black}{ The pattern of the strongest tones is characterised, for all plate-edge streamwise positions, by a `squeezing-together' of the peaks with increasing frequency and Mach number. A second kind of pattern is more clearly visible in the low-Strouhal region (approximately $St<0.6$) of the sound-field PSD maps. This comprises spectral peaks that are both weaker and broader than those at higher frequencies, and in which the `frequency-squeezing' is not so clearly apparent. Our goal is to understand and model the flow dynamics responsible for these two behaviours.}

\section{Understanding and predicting the tones}\label{sec:theory}

Exploration of the mechanisms responsible for the tonal behaviour discussed above requires consideration of the different kinds of wave that are supported by a turbulent jet. This has been done in a pair of recent papers, in locally parallel and weakly non-parallel frameworks by \citet{TowneetalJFM2017}, and in a fully global framework by \citet{SchmidtetalJFM2017}. The studies show that the turbulent jet in isolation can support diverse, weak, forced-resonance mechanisms, due to a rich variety of waves that is briefly discussed in what follows. For a more involved discussion the reader should refer to the said papers.

\subsection{Overview of wave behaviour supported by turbulent jets} 

\begin{figure}
\begin{center}
\includegraphics[trim=0cm 7cm 0cm 7cm, clip, width=0.9\textwidth]{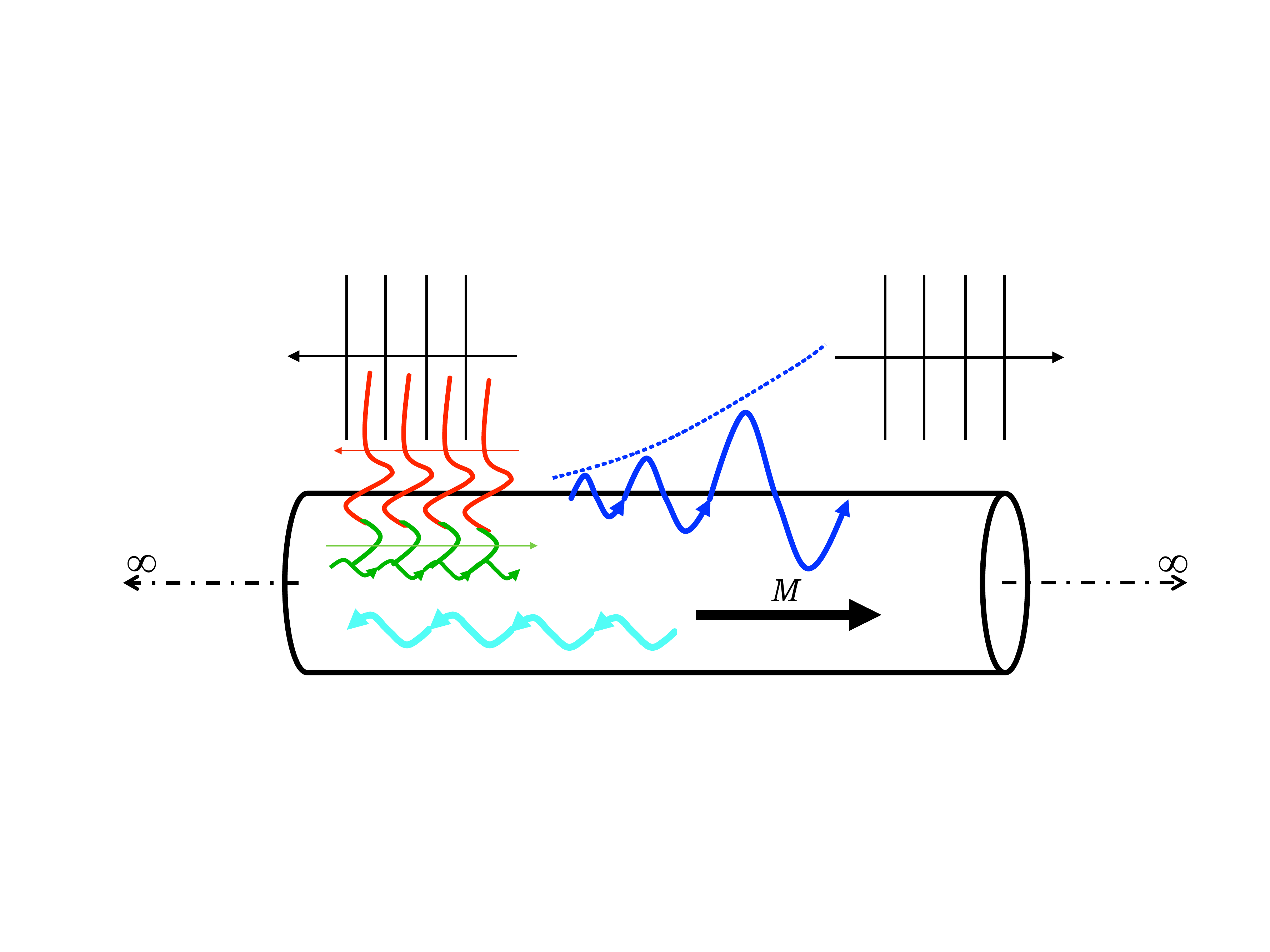}
\caption{Schematic depiction of waves supported by cylindrical vortex sheet; colours correspond to those of figure \ref{fig:DR}.}\label{fig:tube2}
\end{center}
\end{figure}

\begin{figure}
\begin{center}
\includegraphics[trim=8.5cm 17.5cm 8.5cm 17.5cm,clip,width=\textwidth,clip=]{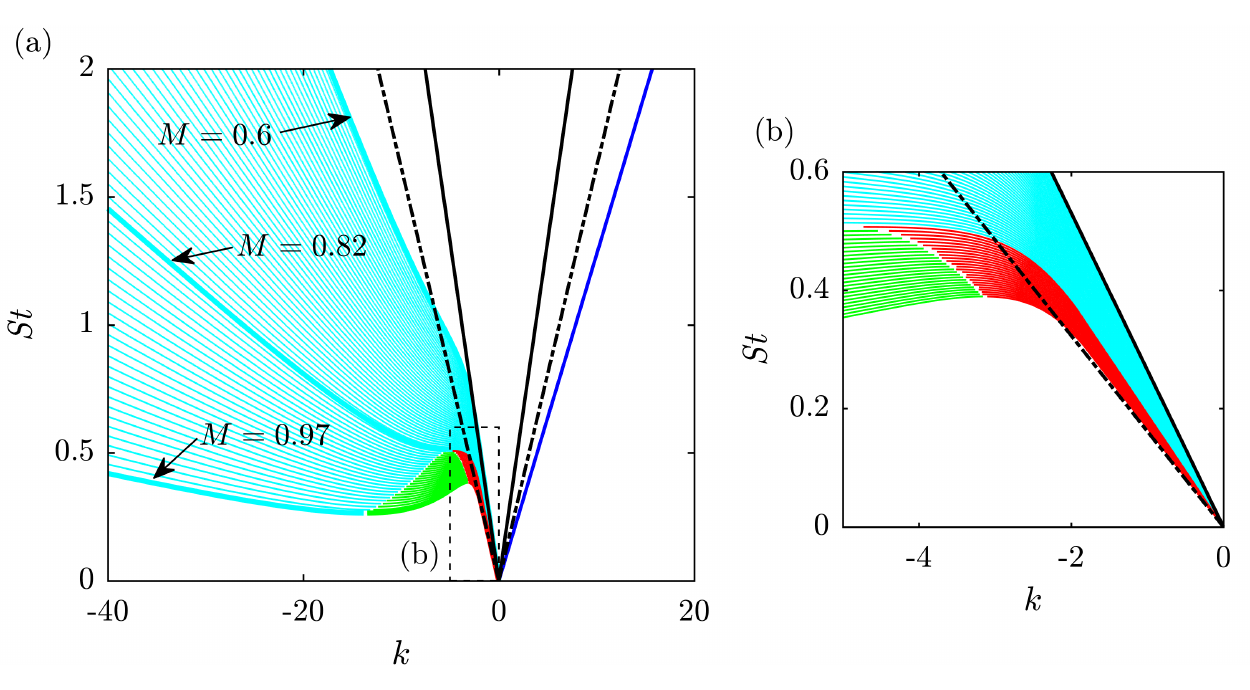} 
\caption{Vortex-sheet dispersion relations in the range $0.6\leq M \leq 0.97$. Blue: $k^+_{KH}$ Kelvin-Helmholtz modes; cyan in range $0.6 \leq M \leq 0.82$: $k^-_{TH}$ modes \citep{tam1989three}; cyan in range $0.82<M\leq0.97$: $k_d^-$ modes \citep{TowneetalJFM2017}; red \& green, respectively: $k^-_p$ and $k^+_T$ acoustic jet modes \citep{TowneetalJFM2017}; black: $k^+$ and $k^-$ freestream sound waves (solid: $M=0.6$; dash-dot: $M=0.97$).}\label{fig:DR}
\end{center}
\end{figure}

The resonance mechanisms studied by \citet{TowneetalJFM2017} and \citet{SchmidtetalJFM2017} in an isolated, isothermal, Mach 0.9 turbulent jet involve two kinds of downstream-travelling ($k^+$) wave: (1) the well-known Kelvin-Helmholtz instability, that we denote $k^+_{KH}$ (blue in figures \ref{fig:tube2} and \ref{fig:DR}); and, (2) \textcolor{black}{a wave discovered by \citet{TowneetalJFM2017}, denoted $k^+_T$ (green in figures \ref{fig:tube2} and \ref{fig:DR}), that only exists in the Mach number range, $0.82 < M \leq 1$, over a restricted range of frequencies, and whose physics vary within that range: at the low-frequency end the waves are largely trapped  within and guided by the jet, behaving in the manner of acoustic waves propagating in a porous-walled cylindrical duct; at the high-frequency end, on the other hand, the waves have support in the shear-layer and it is more appropriate to think of them as shear-layer modes.}

The downstream-travelling waves can undergo resonance with \textcolor{black}{three} kinds of upstream-travelling ($k^-$) wave. One of the upstream-travelling waves is that previously discussed by \citet{tam1989three}, \textcolor{black}{that we denote $k^-_{TH}$} (cyan in figures \ref{fig:tube2} and \ref{fig:DR}). As shown by \citet{TowneetalJFM2017}, \textcolor{black}{this wave exists over the Mach-number range, $0 < M \leq 0.82$, and its physics depend, in a manner similar to that of the $k^+_T$ wave, on the frequency considered: at sufficiently high frequency it is trapped within and guided by the jet, behaving in the manner of a porous-walled acoustic duct mode; at lower frequencies, on the other hand, the mode has support in the shear layer and again must be thought of as a shear-layer mode}. \textcolor{black}{The second $k^-$ wave, also discovered by \citet{TowneetalJFM2017}, exists over the Mach number range, $0.82<M<1$, and has the same soft-duct-like character as the high-frequency $k^-_{TH}$ waves; it is therefore denoted, $k^-_d$ (cyan in figures \ref{fig:tube2} and \ref{fig:DR}). We distinguish it from the $k^-_{TH}$ waves because, unlike these, it becomes evanescent below a well-defined frequency (at the transition from cyan to green in figure \ref{fig:DR}). We note that for Mach numbers close to $M=0.82$, in the close vicinity of the cut-off frequency, it also ceases to be duct-like, but the Strouhal- and Mach-number ranges over which it is duct-like is sufficiently large to justify the denomination. The third upstream-travelling wave behaves in a manner similar to the low-frequency end of the $k^-_{TH}$ branch, i.e. it is a shear-layer mode, and is distinguished from the $k^-_{TH}$ wave by the fact that it becomes evanescent \textit{above} a well-defined frequency (at the transition from red to green in figure \ref{fig:DR}). This wave is denoted $k^-_p$ (red in figures \ref{fig:tube2} and \ref{fig:DR})}. We can add to this catalogue of waves, upstream- and downstream-travelling freestream sound waves (black in figures \ref{fig:tube2} and \ref{fig:DR}), which are also potential candidates for the enabling of resonance. 

\textcolor{black}{The $k_{TH}^-$, $k_T^+$, $k_d^-$ and $k_p^-$ waves are members of hierarchical families parameterised by two integers, $(m,j)$, corresponding to the azimuthal and radial orders of the waves. In what follows, we restrict attention to axisymmetric waves, $m=0$, of radial order, $j=1$. }

\subsection{Dispersion relations and preliminary tone-frequency prediction}

Our objective is to see if the tone patterns can be understood in terms of the waves described above. \textcolor{black}{The linearised Euler equations provide the modelling framework. These are considered in a locally parallel setting, with normal-mode \textit{Ansatz},
\begin{align}
 q(x,r,t)=\hat{q}(r)\text{e}^{i(kx-\omega t)}.
 \end{align} 
 Here, $k$ is the streamwise wavenumber, non-dimensionalised by $D$, and, $\omega=2\pi St M$, is the non-dimensional frequency.} Three dispersion relations, corresponding to increasingly realistic conditions, are obtained from these: (DR1) that which is obtained by considering the jet to behave as a porous-walled cylindrical duct; (DR2) that which describes waves supported by a cylindrical vortex sheet \citep{lessen1965inviscid,michalke1970note}; and, (DR3) that which is obtained if the shear-layer is considered to have finite thickness. \textcolor{black}{These three models have been thoroughly discussed by \cite{TowneetalJFM2017}.} We make a preliminary tone-prediction using DR2; fine-tuning requires consideration of DR1 and DR3.

Figure \ref{fig:DR} shows vortex-sheet dispersion relations, DR2, in the Mach-number range $0.6\leq M\leq 0.97$. With the exception of the Kelvin-Helmholtz mode, which has non-zero imaginary part, the lines are locii of eigenvalues with zero imaginary part, i.e. neutrally stable, propagating waves \citep{TowneetalJFM2017}. The waves discussed above have been colour-coded. The downstream-travelling waves are shown in blue and green: respectively, the Kelvin-Helmholtz mode, $k^+_{KH}$, and the $k_T^+$  mode. Upstream-travelling waves are shown in cyan and red: respectively, $[k^-_{TH}$ ($0.6 \leq M \leq 0.82 $); $k^-_d$ ($0.82< M<1$)] and $k^-_p$ ($0.82 < M<1$). The black lines show dispersion relations for upstream- and downstream-travelling free-stream sound waves at $M=0.6$ (solid) and, $M=0.97$ (dash-dot).

\begin{figure}
\begin{center}
\includegraphics[trim=8.35cm 17.5cm 8.5cm 17.5cm,width=\textwidth,clip=]{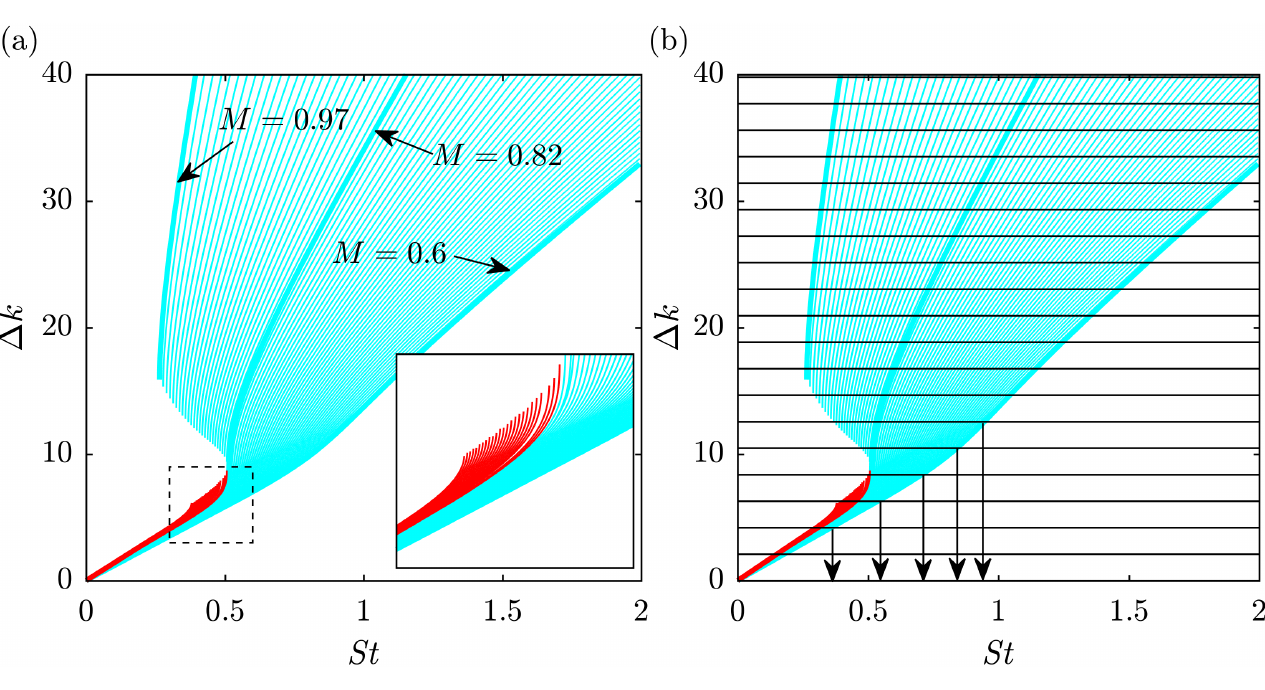} 
\caption{(a) $\Delta k$ between $k_{KH}^+$ Kelvin-Helmholtz mode and all $k^-$ jet modes in range $0.6\leq M \leq 0.97$; (b) Illustration of resonance-frequency identification (showing `frequency squeezing')  for $L/D=3$, $M=0.6$ and \textit{hard-hard} end conditions: horizonal lines show values of $\Delta k_{hh}$ (equation \ref{eq:dkss}) for L/D=3.}\label{fig:DK}
\end{center}
\end{figure}

Resonance can potentially occur between any $k^+/k^-$ mode pair; there are therefore five different possibilities if we exclude resonance between upstream- and downstream-travelling sound waves. Given that the Kelvin-Helmholtz mode is the only unstable wave, all others being in reality  either neutral or slightly damped, the most likely scenario is that in which $k^+_{KH}$ is coupled, via end conditions provided by the nozzle exit plane and the plate edge, to a $k^-$ mode. \textcolor{black}{A further argument for excluding the $k^+_T$ wave is the continuity of the tones across the $M=0.82$ threshold, on the lower side of which these modes are evanescent.}

We consider two possible pairs of end conditions: 
a \textit{hard-hard} condition, $\hat{u}(x=0,\omega)=\hat{u}(x=L,\omega)=0$, in which upstream-travelling and downstream-travelling waves are reflected by solid walls (a simplified model for scattering by the nozzle lip and the plate edge); and a \textit{soft-hard} condition, $\hat{p}(x=0,\omega)=\hat{u}(x=L,\omega)=0$, in which the upstream-travelling waves are reflected by pressure-release in the nozzle exit plane, downstream-travelling waves being reflected by the plate edge, again modelled as a solid wall.

\begin{figure}
\begin{center}
{\includegraphics[trim=0 0 0 0, clip, width=1\textwidth]{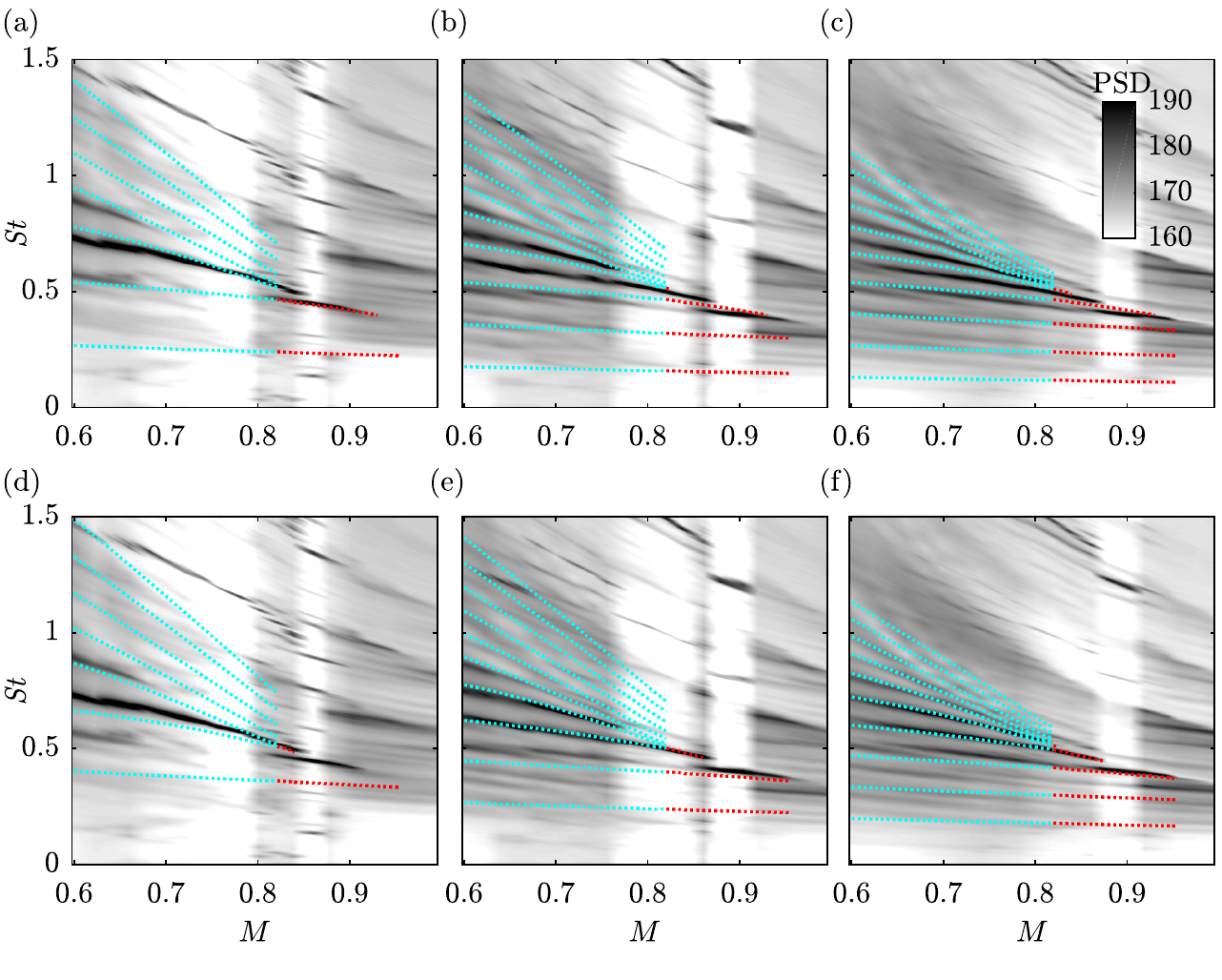}}
\caption{Tone-frequency predictions using vortex-sheet dispersion relations and assuming resonance between $k_{KH}^+$ and $k^-$ jet modes, coupled by \textit{hard-hard} (top) and \textit{soft-hard} (bottom) end conditions. From left to right: $L/D=2$, $3$ \& $4$. Cyan: resonance between $k_{KH}^+$ and $k_{TH}^-$; red: resonance between $k_{KH}^+$ and $k_p^-$.}\label{fig:comp_oo_oc}
\end{center}
\end{figure}

It is straightforward to show that these scenarios lead to the following resonance criteria,
\begin{align}
\Delta k_{hh}=&\frac{2n\pi}{L}, \label{eq:dkss}\\
\Delta k_{sh}=&\frac{(2n+1)\pi}{L},\label{eq:dkhs}
\end{align}
for \textit{hard-hard} (\textcolor{black}{note that \textit{soft-soft} end conditions would lead to the same criterion)} and \textit{soft-hard} end conditions, respectively, where $\Delta k$ is the difference in wavenumber between the upstream- and downstream-travelling waves that participate in the resonance. 

The difference in wavenumber, $\Delta k$, can be easily computed for any $k^+/k^-$ pair as a function of Mach number and frequency, using the dispersion relations of the two waves. This is shown in figure \ref{fig:DK}(a) for the pairs $k^+_{KH}/k^-_{TH}$ and $k^+_{KH}/k^-_d$, both shown in cyan, and for $k^+_{KH}/k^-_p$, shown in red. Having calculated $\Delta k$, the resonance criteria of equations \ref{eq:dkss} and \ref{eq:dkhs} can be superposed, as in figure \ref{fig:DK}(b), and the resonance frequencies provided by the intersection of these with the lines $\Delta k(M,St)$. The example shown in figure \ref{fig:DK}(b) is for $M=0.6$, $L/D=3$ and \textit{hard-hard} end conditions, and it illustrates an interesting characteristic of this kind of resonance: a frequency `squeezing', due to the dispersive nature of the $k^-$ waves. As the Mach number is increased this `squeezing' becomes more pronounced, due to the stronger variation of phase speed with frequency.

Tone-frequency predictions are made for the Mach-number range considered, using both \textit{hard-hard} and \textit{soft-hard} end conditions, and these are compared with the observed behaviour in figure \ref{fig:comp_oo_oc}. The general trend is found to be satisfactorily captured, in particular the aforesaid frequency `squeezing'. But there remain three discrepancies: (1) the resonance models predict a continuation of the tones to infinitely high frequency, whereas the data clearly shows a frequency cut-off; (2) at low frequencies, the model predicts peaks that are not observed in the data; (3) the best match is for some cases provided by the \textit{hard-hard} end conditions, whereas for others the \textit{soft-hard} conditions do better.

The third of these discrepancies may be an indication that the flow is selecting between different end conditions, and indeed an observation of mode switching---clearly audible in the experiments---provides support for this idea. The discrepancy may, alternatively, be due to the actual end conditions being more subtle than $\hat{u}=0$ and $\hat{p}=0$, comprising flow physics only amenable by a Wiener-Hopf analysis \textcolor{black}{involving boundary conditions that change at $x=0$ between a hard-walled nozzle and a vortex sheet}, or by detailed numerical analysis. Both are beyond the scope of the present work, and we will therefore not explore this point further, accepting that the end-condition modelling, whilst not perfect, is sufficient to support the main conclusion of the work, which is that the $k^-$ jet modes of \citet{TowneetalJFM2017} and \citet{SchmidtetalJFM2017}, \textcolor{black}{together with Kelvin-Helmholtz wavepackets and free-stream sound waves}, underpin the observed edge-tone behaviour. Where the two other discrepancies are concerned, on the other hand, further explanation is possible. The high-frequeny cut-off is considered in what follows.

\subsection{High-frequency tone cut-off}

In the Mach-number range $0.6\leq M\leq 0.82$ we explore the high-frequency cut off using DR3 and DR1. The first illustrates a progressive trapping of the $k^-_{TH}$ wave by the jet with increasing frequency ; once entirely trapped (duct-like) it is prevented from interacting with the nozzle lip. The second is used to show that for frequencies at which the $k^-_{TH}$ wave is truly duct-like (trapped), it cannot be reflected by pressure release in the nozzle exit plane and is entirely transmitted into the nozzle. In the Mach-number range, $0.82\leq M<1$, on the other hand, the resonance cut-off condition is due simply to a cutting-off of the $k^-_p$ waves: \textcolor{black}{as discussed earlier,} they are evanescent above a well-defined frequency. \textcolor{black}{This can be seen in figure \ref{fig:DR} by looking at the red lines, which only appear up to a given Strouhal number; beyond this value the waves become evanescent, which corresponds to the saddle-point $S2$ discussed in \cite{TowneetalJFM2017}.} 


\subsubsection{Trapped waves cannot touch the nozzle}

\begin{figure}
\begin{center}
{\includegraphics[trim=4.5cm 11.5cm 4.7cm 11.5cm, clip, width=0.8\textwidth]{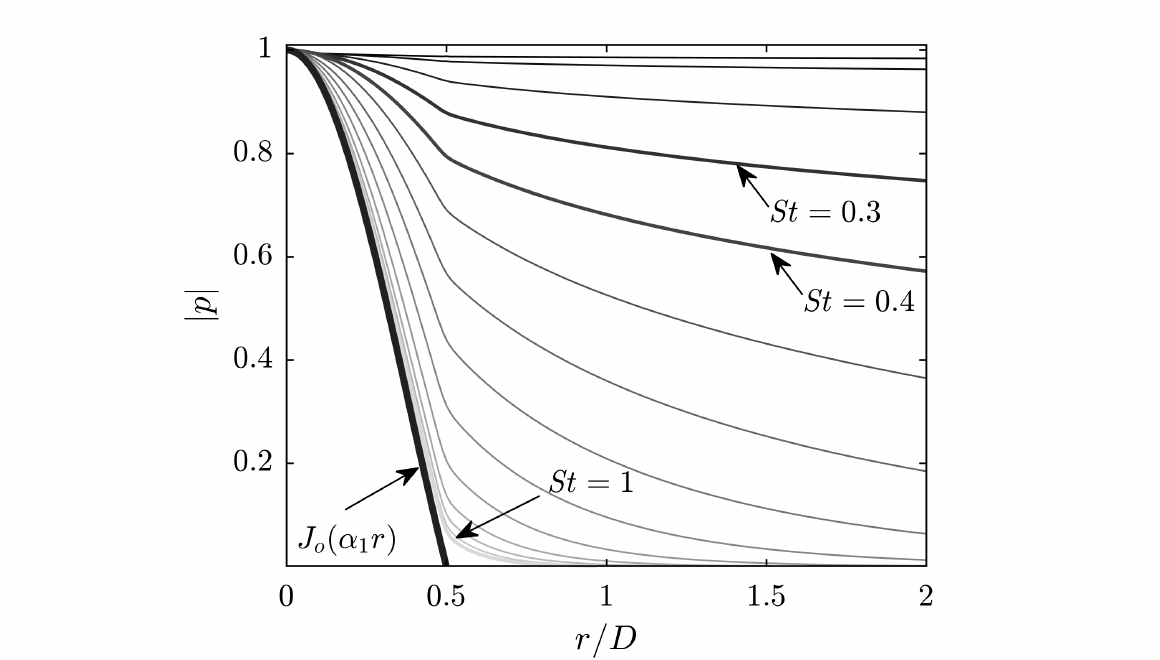}}
\caption{Pressure eigenfunctions associated with $k_{TH}^-$ jet modes in $M=0.6$ finite-thickness round jet, in frequency range $0.05\leq St \leq 1$. The Bessel function, $J_o(\alpha_1r)$, is a porous-walled-duct mode of radial order $1$.}\label{fig:eigfunc}
\end{center}
\end{figure}

To understand the high-frequency cut-off in the range $0.6\leq M\leq 0.82$ we first consider the dispersion relation associated with a finite-thickness round jet, $DR3$. Eigensolutions of the Rayleigh equation are obtained for an axisymmetric mixing layer whose radial velocity profile has the form of a hyperbolic-tangent velocity profile with displacement thickness similar to that of the $M=0.6$ free jet at $x/D=1$. 


We are here interested in the frequency dependence of the pressure eigenfunctions associated with the discrete mode equivalent to the $k_{TH}^-$ mode in the vortex-sheet model. This is shown in figure  \ref{fig:eigfunc}. At very low frequencies the waves behave much like plane free-stream sound waves; the scale separation between their wavelength and the width of the jet is such that they are essentially unaffected by the flow. With increasing frequency they become progressively deformed/trapped. At $St\approx 1$ their radial support shows them to be almost entirely confined within the jet. As shown by \cite{TowneetalJFM2017}, they here behave in the manner of waves propagating in a soft-walled duct, experiencing the shear-layer as a pressure-release surface. This can be seen here in the similarity of the high-frequency, finite-thickness eigenfunctions to those of soft-duct waves, which take the form of Bessel functions, $J_o(\alpha_j r)$. The implication for resonance is that at these high frequencies, $k^-_{TH}$ waves impinging on the nozzle exit plane have negligible fluctuation levels in the radial vicinity of the nozzle lip, and are thus deprived of the possibility of  there being scattered into $k^+$ waves. One of the possible reflection mechanisms necessary to sustain resonance is thereby disabled.

\subsubsection{Disabled pressure release in range $M\leq0.82$}

We have seen that  pressure-release in the nozzle exit plane, the `soft' end condition, may also provide a reflection mechanism, particularly so as the $k^-_{TH}$ waves become more duct-like at these higher frequencies.  We therefore consider the nozzle-jet system as a rigid-walled cylindrical duct connected to a porous-walled duct, the ensemble containing a plug flow of Mach number, $M\leq0.82$. As discussed in the Appendix, for the frequencies of interest, 
an incident $k^-_{TH}$ wave of amplitude $I$ and radial order $j=1$ is transmitted as a plane wave, and reflected as a $k^+$ porous-walled duct mode of amplitude $R$ and radial order $j=1$. The absolute value of the reflection coefficient, 
\begin{equation}
\frac{R}{I}=\Big(\frac{1-\frac{k_j^+M}{\omega}}{1-\frac{k_j^-M}{\omega}}\Big)\Big(\frac{k_n^--k_j^-}{k_j^+-k_n^-}\Big),\label{eq:RoI}
\end{equation}
derived in the appendix, 
is shown in figure \ref{fig:RoI} as a function of Mach and Strouhal numbers. 

\begin{figure}
\begin{center}
{\includegraphics[trim=0 0 0 0, clip, width=1\textwidth]{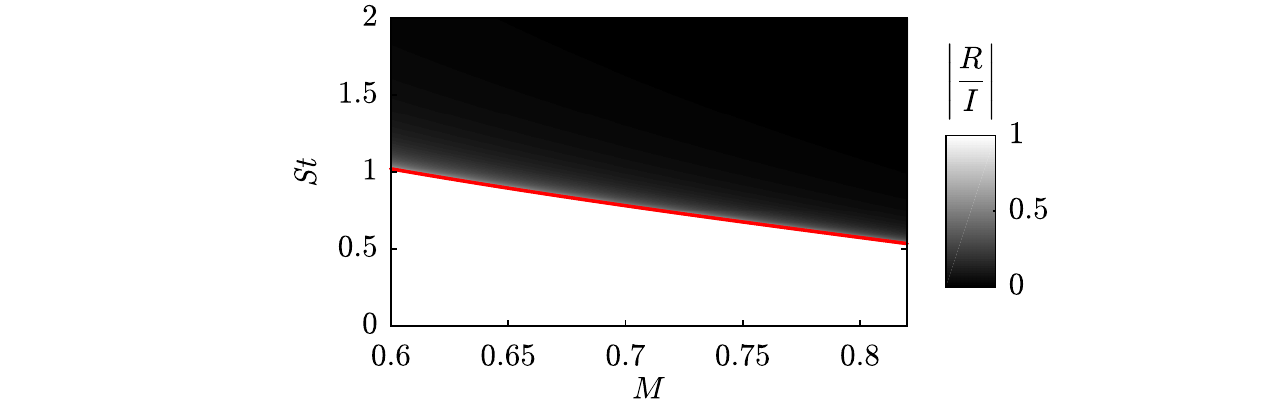}}
\caption{Reflection coefficient, $|\frac{R}{I}|$, for a $k^-$, \textit{porous-walled-duct} mode impinging on nozzle plane. The red line shows the cut-on frequency. At frequencies below this line, the incident wave is evanescent; above it is propagative, but has zero reflection coefficient: it is entirely transmitted into the nozzle as a rigid-walled-duct $k^-$ plane wave.}\label{fig:RoI}
\end{center}
\end{figure}

Note that the red line corresponds to the cut-on condition for porous-walled duct modes: for frequencies below this line such modes are evanescent. But as we have already seen, the porous-walled duct is not a good model for $k^-$ jet modes in the low-frequency range; these are propagative in both the vortex-sheet and finite-thickness mixing-layer models, and have significant radial support across the shear layer. Above the porous-duct cut-on frequency, on the other hand, where the $k^-_{TH}$ jet modes have become trapped, as the eigenfunctions in figure \ref{fig:eigfunc} show, the porous-walled duct is a good approximation for the dynamics of these waves. We are therefore only interested in the reflection coefficient of equation \ref{eq:RoI} above this cut-on frequency, where, as shown in figure \ref{fig:RoI}, $|\frac{R}{I}|\approx 0$. 

\begin{figure}
\begin{center}
{\includegraphics[trim=10cm 17.5cm 10cm 17.5cm, clip, width=0.85\textwidth]{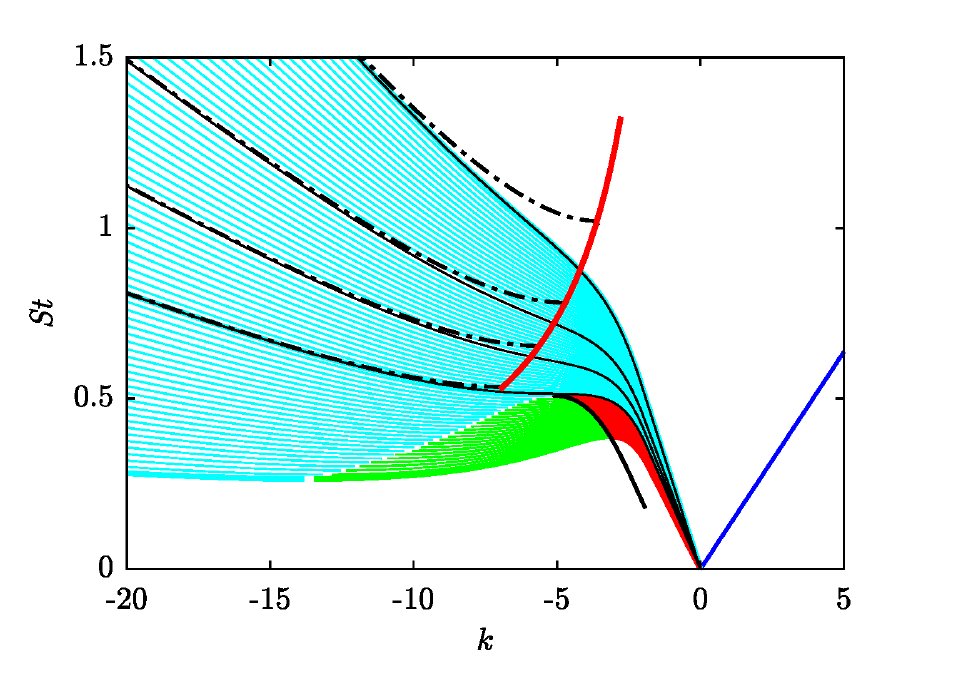}}
\caption{Truncation of dispersion relations: (1) To account for trapping of $k^-_{TH}$ waves, which occurs above the thick solid red line, the cut-on condition for porous-walled duct modes; and, (2) To account for saddle-point cut-off of $k^-_p$ waves, indicated by the solid black line. Final tone-frequency predictions, shown in figure \ref{fig:comp_final_oo_oc} are made using eigenvalues to the right of the solid red and black lines. The dash-dotted black lines show dispersion relations for a porous-walled duct for four Mach numbers ($M=0.6$, $0.7$, $0.75$ \& $0.82$), allowing comparison with the vortex-sheet dispersion relations.}\label{fig:DR_truncated}
\end{center}
\end{figure}

\begin{figure}
\begin{center}
{\includegraphics[trim=0 0 0 0, clip, width=1\textwidth]{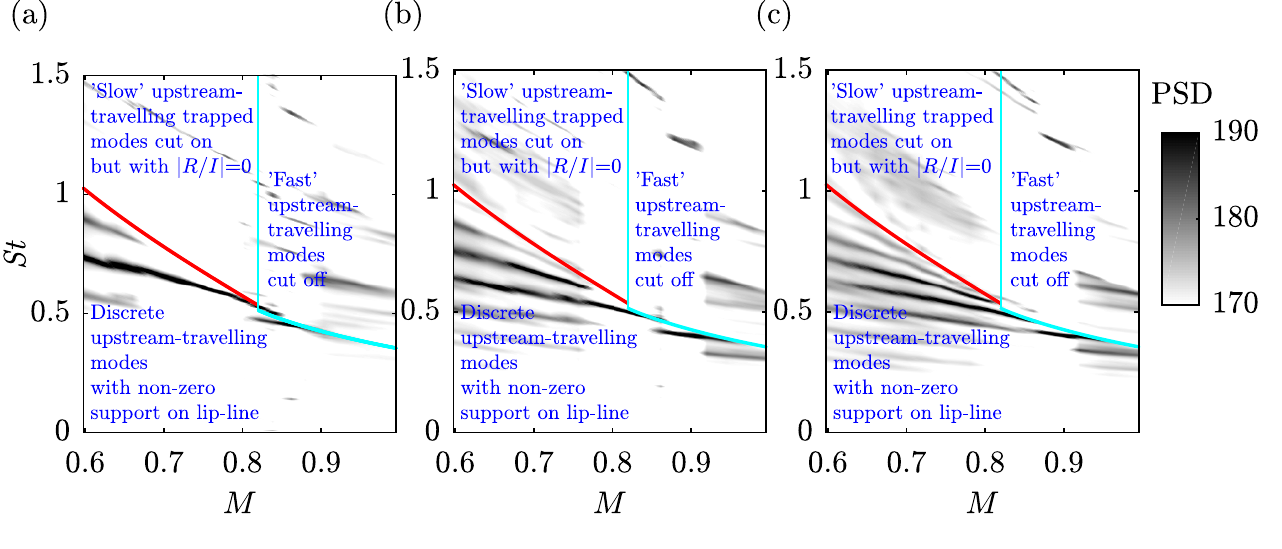}}
\caption{Comparison of hypothesised resonance cut-off mechanims with data. Red line: resonance cut-off due to trapping of $k^-_{TH}$ waves, which prevents interaction with nozzle lip, and leads to the disabling of pressure release in the nozzle exit plane; thick cyan line: resonance cut off due to saddle-point cut off of $k^-_p$ waves.}\label{fig:cutoff_comp_exp}
\end{center}
\end{figure}

\begin{figure}
\begin{center}
{\includegraphics[trim=0 0 0 0, clip, width=1\textwidth]{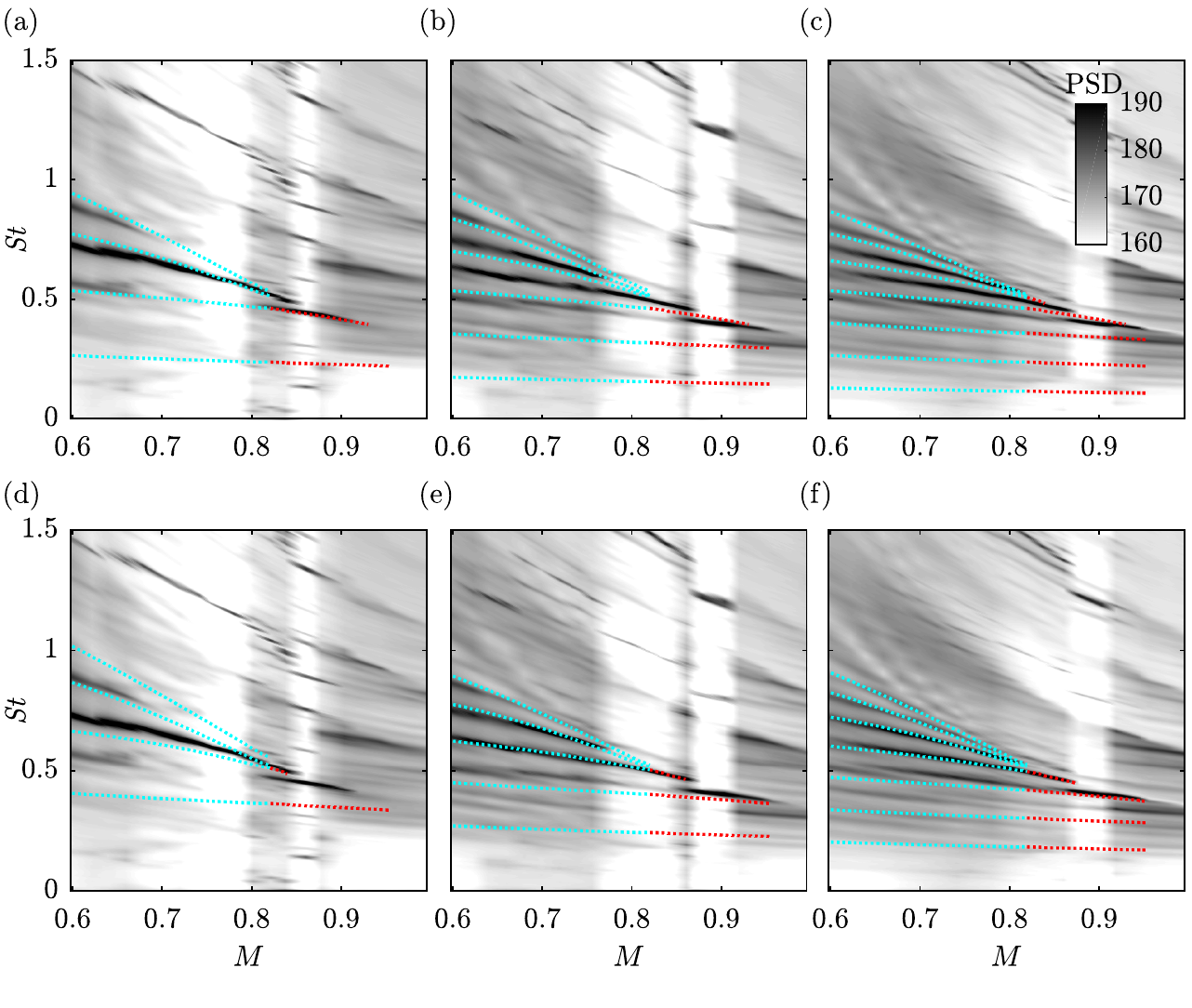}}
\caption{Refined tone-frequency predictions: using vortex-sheet dispersion relations, assuming resonance between Kelvin-Helmhotz $k^+$ and $k^-$ jet modes, coupled by \textit{hard-hard} (top) and \textit{hard-soft} (bottom) end conditions; assuming high-frequency cut-off due to trapped, non-reflecting $k_{TH}^-$ jet modes in the range $0.6 \leq M \leq 0.82$, and due to cut off $k_p^-$ jet modes in the range $0.82 < M \leq1$. From left to right: $L/D=2$, $3$ \& $4$.}\label{fig:comp_final_oo_oc}
\end{center}
\end{figure}

The $k^-_{TH}$ waves above this frequency impinging on the nozzle exit plane are thus not reflected by the pressure-release condition, they are  almost entirely transmitted into the nozzle where they propagate as plane waves. This result, combined with that of the previous section that shows how trapping of the $k^-_{TH}$ waves leads to the absence of significant fluctuation levels in the vicinity of the nozzle lip, demonstrates how, for $M\leq 0.82$, both the \textit{hard} and \textit{soft} reflection conditions are disabled above the porous-walled-duct cut-on frequency. 

This frequency is superposed on the vortex-sheet dispersion relations in figure \ref{fig:DR_truncated} (solid red line), demarcating regions where the DR2 (coloured lines) and DR1 (black dash-dotted line, shown for four Mach numbers) become similar (above the red line) and where they are not (below). In the region above this line the $k^-_{TH}$ vortex-sheet and finite-thickness mixing-layer modes behave essentially as propagative porous-walled duct modes, trapped such that they avoid the nozzle lip, and with zero pressure-release reflection in the nozzle exit plane. Resonance will thus be disabled for frequencies above this line, which is superposed on the data in figure \ref{fig:cutoff_comp_exp} and found to agree well with the observed resonance cut-off.
\subsubsection{Saddle-point cut-off of $k^-_p$ modes in range $0.82< M<1$}

To complete the picture we must consider the resonance cut-off condition in the Mach number range $0.82 < M<1$, where the $k^-_p$ modes, which never behave like porous-walled duct modes, are those that underpin resonance. This is straightforward: as shown by the solid black line in figure \ref{fig:DR_truncated} these modes become evanescent above a certain Strouhal number. The cut-off Strouhal number is plotted on the data in figure \ref{fig:cutoff_comp_exp} as a thick cyan line, where we see that it corresponds well with the observed resonance cut-off in this Mach-number range. 

\subsection{Refined tone-frequency prediction}

\begin{figure}
\begin{center}
{\includegraphics[trim=0 1cm 0 0.5cm,clip,width=1\textwidth]{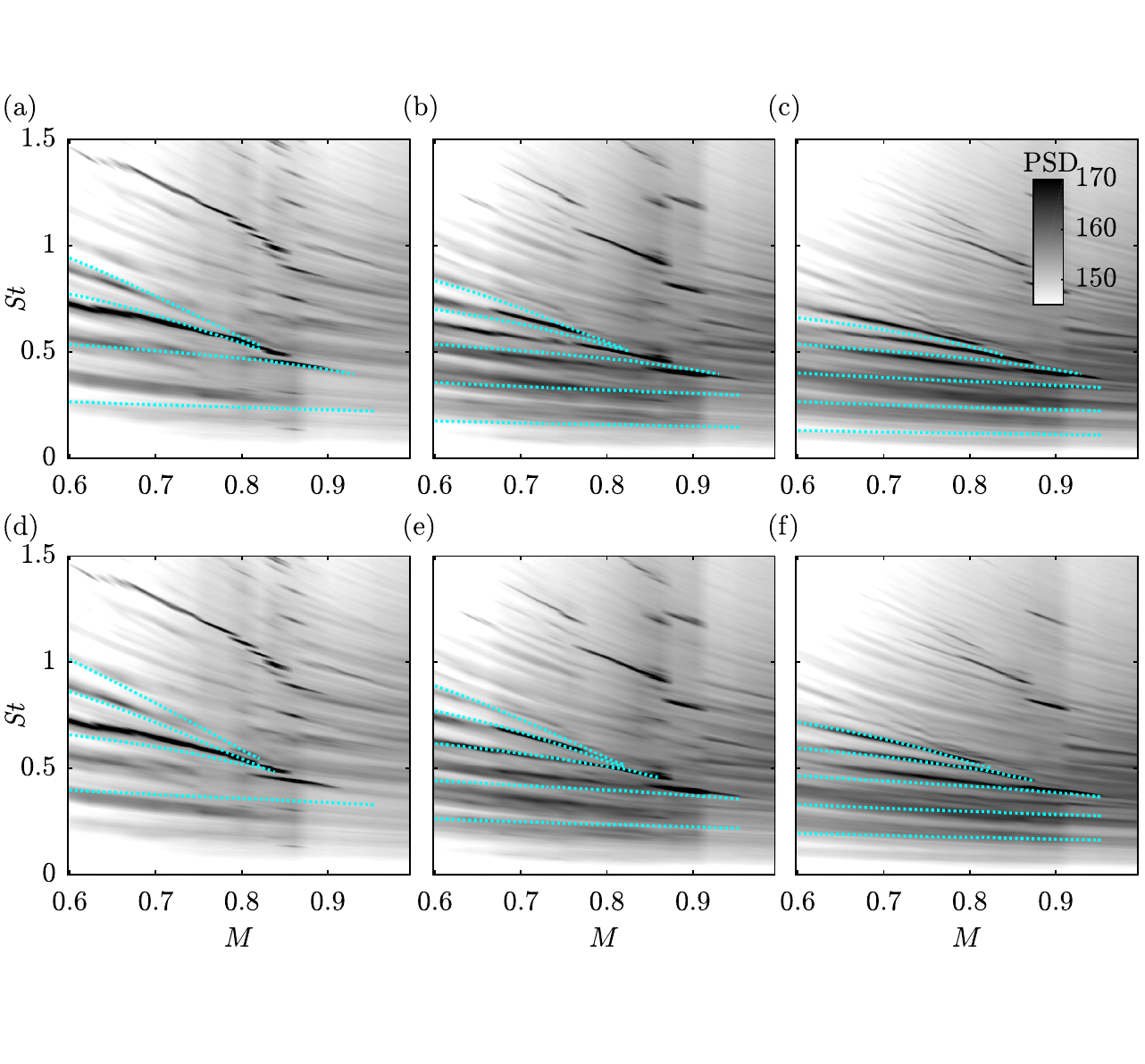}}
\caption{Acoustic tone-frequency predictions assuming $k_{KH}^-$ resonance. Top: `hard-hard' boundary conditions; bottom: `soft-hard' boundary conditions. From left to right: $L/D=2$, $3$ \& $4$.}\label{fig:comp_kkh}
\end{center}
\end{figure}

Tone-frequency predictions can now be made using the vortex-sheet dispersion relation, truncated such that  only eigenvalues to the right of the solid red and black lines in figure \ref{fig:DR_truncated} are used. The refined tone predictions are shown in figure \ref{fig:comp_final_oo_oc}. 

{\textcolor{black}{Recall that throughout we have been exclusively modelling axisymmetric waves of radial order 1, which are responsable for the lowest-frequency band of peaks. The higher-frequency bands, which are generally clearest for $M>0.9$, are due to higher radial and azimuthal wave orders that we do not attempt to model because they do not show up with sufficient clarity in the single-point microphone measurements we consider.}}  

The match between prediction and observation can be considered satisfactory in the $St-M$ range where strong resonance is observed, given the aforesaid caveat regarding the simplicity of the end-condition modelling. But there remain discrepancies at low frequency, where a match is not clear between model and data. This is considered in what follows.


\subsection{What is happening at low frequency?}

\begin{figure}
\begin{center}
{\includegraphics[trim=0 1cm 0 0.5cm,clip,width=1\textwidth]{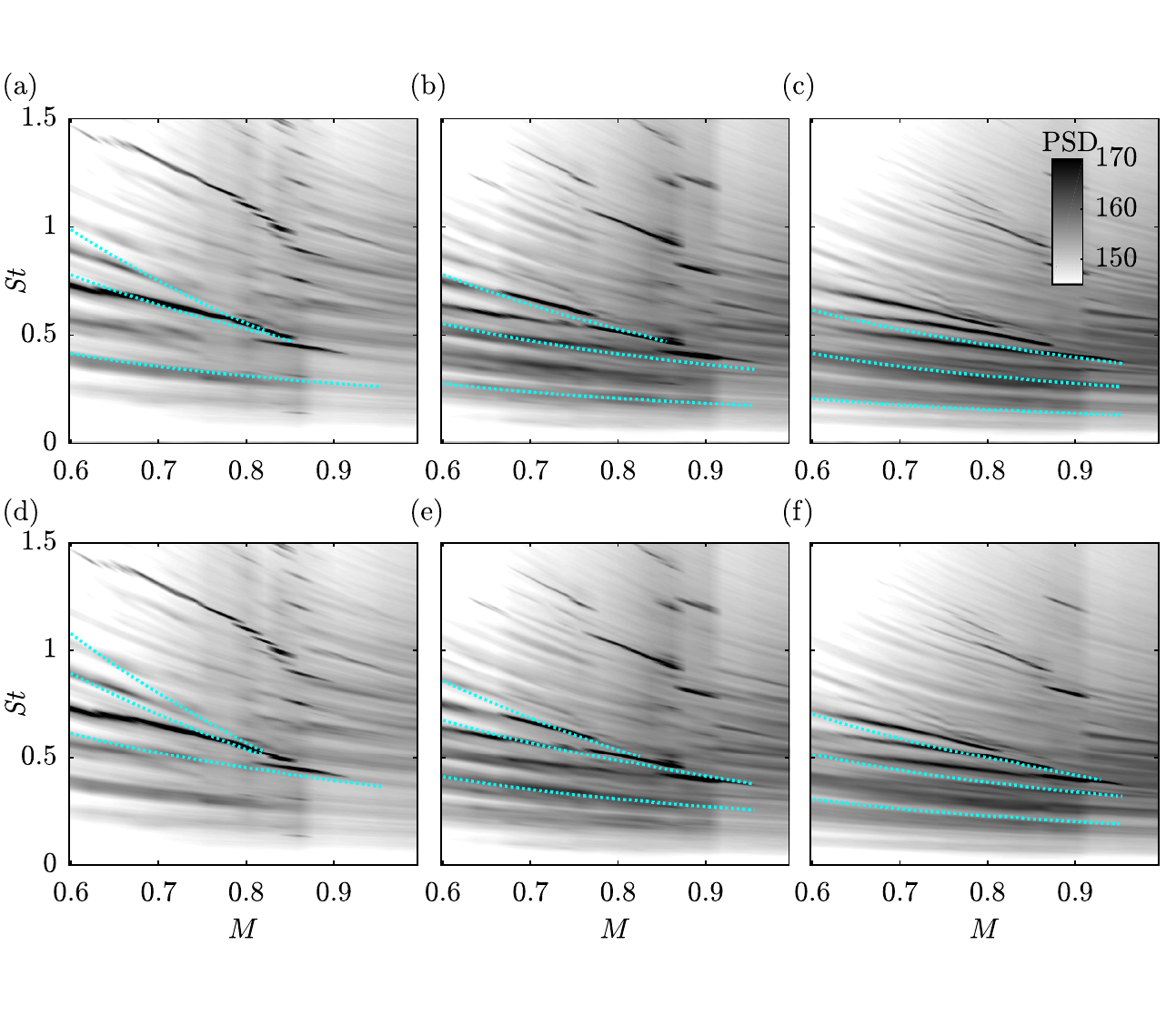}}
\caption{Acoustic tone-frequency predictions assuming $k_a^-$ resonance (between $k^-$ jet modes and free-stream $k^+$ sound waves). Top: `hard-hard' boundary conditions; bottom: `soft-hard' boundary conditions. From left to right: $L/D=2$, $3$ \& $4$.}\label{fig:comp_ka}
\end{center}
\end{figure}

\textcolor{black}{In order to understand the low-frequency regime we reconsider the PSD maps of figure \ref{fig:acou}, which were generated using data from the acoustic field. As discussed earlier, these reveal a low-frequency pattern of spectral peaks not so clearly visible in the nearfield. These are both weaker and broader than those at higher frequency; and they show a different $St-M$ dependence, not matched by predictions based on resonance between $k_{KH}^+$ and the $k^-$ waves considered (we will refer to this as `$k_{KH}$ resonance' in what follows), as shown by figure \ref{fig:comp_kkh}. Note that we have now dropped the colour-coding, and it is to be understood  that predictions in the ranges $0.6\leq M\leq0.82$ and $0.82<M\leq 0.97$ are made using, respectively, the $k_{TH}^-$ and $k_p^-$ waves.}

\textcolor{black}{To model the low-frequency tones, we consider a different resonance scenario involving upstream-traveling $k_{p}^{-}$ and $k_{TH}^{-}$ waves (for $M>0.82$ and $M\leq 0.82$, respectively) and downstream-travelling sound waves. We refer to this scenario, in what follows, as `$k_a$ resonance'. The tones predicted by this scenario are shown in figure \ref{fig:comp_ka}. In the $St-M$ range where the $k_{KH}$ mechanism has already been shown to underpin the observed tones, the $k_a$ model does not match the data, as expected. At lower frequency, on the other hand, we see that if the upstream reflection mechanism is considered to be dominated by the nozzle lip, the $n=1$ ($n$, the resonance order, is defined in equations \ref{eq:dkss} and \ref{eq:dkhs}), $k_a$ prediction matches the first peaks with good accuracy for edge positions, $L/D=2$ and $3$, as seen in figure \ref{fig:comp_ka} (a) \& (b); for $L/D=4$, shown in figure \ref{fig:comp_ka} (c), there is slight mismatch in terms of the precise value of the resonance frequencies, but the slope is accurately reproduced. The $n=2$ prediction matches the low-Mach trend of the \textit{third} observed tone for $L/D=3$, as seen in figure \ref{fig:comp_ka} (b).  If, on the other hand, reflection at the upstream boundary is considered to be dominated by pressure release in the nozzle exit plane, the $n=1$, $k_a$ prediction matches the second peak for $L/D=3$, as shown in figure \ref{fig:comp_ka} (e), and the low-Mach trend of the second peak for $L/D=2$, as is clear from figure \ref{fig:comp_ka} (d). For edge-position, $L/D=4$, again there are mismatches in terms of the precise frequencies predicted, but the slopes are correctly modelled. In summary: if two upstream boundary conditions are admitted, both the slope and spacing of the low-frequency tones, and in most cases the precise values of frequency, can be reproduced by the $k_a$ model.}

%
%
%

\textcolor{black}{In a linear framework there can, of course, be only one upstream boundary condition. If this is cast in terms of an acoustic impedance relating pressure and velocity fluctuations at the nozzle exit, a possible explanation for the foregoing observations is that such frequency dependent impedance tends to high absolute value for low frequencies (leading to a hard-wall boundary condition) and to low absolute values as the frequency is increased (resulting in a soft-wall boundary condition). On the other hand, if we admit the presence of non-linear effects in the resonance dynamics---which is certainly the case, as attested to by the harmonics of the strongest tone that can be seen in all of the PSD maps---we may then postulate that the resonance system involves competition between different loop-closing scenarios. This would result in mode switching, where the flow locks onto a $k^+/k^-$ limit cycle that dominates for a certain time before non-linearity bumps it onto another $k^+/k^-$ combination. The flow may similarly be choosing between, or simultaneously exploiting, two upstream boundary conditions: the nozzle lip, which presents an unfailing $u=0$ at $r/D=0.5$ to upstream-travelling waves, and a nozzle-plane impedance, where the same waves experience the condition $p\approx 0$, but which fluctuates. These interesting possibilities should be explored in future work.}

\textcolor{black}{We close the analysis and discussion with some final comments regarding the low- and high-frequency resonance regimes. A first observation concerns the different nature of the peaks in each. In the $k_a^+$ regime these are weak and have a low quality factor, whereas resonance in the $k_{KH}^+$ regime is characterised by strong peaks of high quality factor. This is consistent with the models. The $k_{KH}^+$ resonance involves an unstable downstream-travelling wave coupled with a neutrally stable upstream-travelling wave. An oscillator-like flow behaviour results, and indeed non-linear saturation is observed as discussed above. The $k_a^+$ resonance, on the other hand, involves the coupling of two neutrally stable waves, similar to the resonance observed in free-jet conditions \citep{TowneetalJFM2017,SchmidtetalJFM2017}, and, as is the case with that phenomenon, the mechanism requires forcing by other flow dynamics.}

\textcolor{black}{Finally, why does the flow not exploit the unstable $k_{KH}^+$ modes at low frequency? This may be due to the non-modal nature of low frequency $k^+$ wavepacket dynamics, recently the object of a number of investigations \citep{towne2015stochastic,semeraroAIAA,jeun2016PoF,tissot2016sensitivity,tissot2017orr,jordan2017modal}. As shown by those studies, at low Strouhal numbers, typically for $St<0.4$, $k^+$ dynamics are not dominated by the single, unstable, Kelvin-Helmholtz mode, but involve an ensemble of stable $k^+$ waves, of differing phase speeds, overall downstream growth being underpinned by the non-normal combination of these modes under the influence of stochastic forcing by turbulence. We can postulate that this would inhibit two-mode resonance, and that the flow may `prefer' in that case the simpler $k_a$ resonance mechanism.}

\vspace{-0.5cm}

\section{Conclusion}\label{sec:conclusion}

\textcolor{black}{Experiments have been performed involving an isothermal, round, turbulent jet, whose Mach number is varied, and that grazes a sharp edge whose streamwise position is varied. This leads to rich tonal dynamics and sound that cannot be explained by appealing to the usual edge-tone scenario. The observed behaviour has been analysed by the combined use of three locally parallel linear models. Vortex-sheet dispersion relations provide phase-speed information sufficient to show that the strongest peaks are underpinned by resonance between downstream-travelling Kelvin-Helmholtz wavepackets and two kinds of upstream-travelling jet modes: those considered by \cite{tam1989three} and those of \cite{TowneetalJFM2017} and \citet{SchmidtetalJFM2017}. This model also explains a high-frequency resonance cut-off that occurs in the range, $0.82<M<1$. Resonance cut-off in the range $M\leq0.82$, on the other hand, requires consideration of two further effects. On one hand the upstream-travelling waves are shown to become progressively trapped with increasing frequency, leading to the disabling of their scattering by the nozzle lip. A porous-walled-duct model is then used to show that the upstream-travelling waves, once  trapped, are no longer reflected in the nozzle exit plane by pressure release. Finally, it is shown that at low frequency a weaker, forced resonance exists that involves a coupling between the upstream travelling jet modes and downstream travelling sound waves. This resonance regime is hypothesised to arise on account of the non-modal nature of hydrodynamic $k^+$ wavepackets at low frequency.}\\

\noindent \textbf{{Acknowledgements.}} This work was supported by the EU project JERONIMO (call reference number: FP7-AAT-2012-RTD-1; grant agreement number: 314692).

\newpage

\appendix
\section{\label{appendixB:RC} Nozzle-plane reflection conditions}

\begin{figure}
\begin{center}
{\includegraphics[trim=0cm 10cm 0cm 7cm, clip, width=1\textwidth]{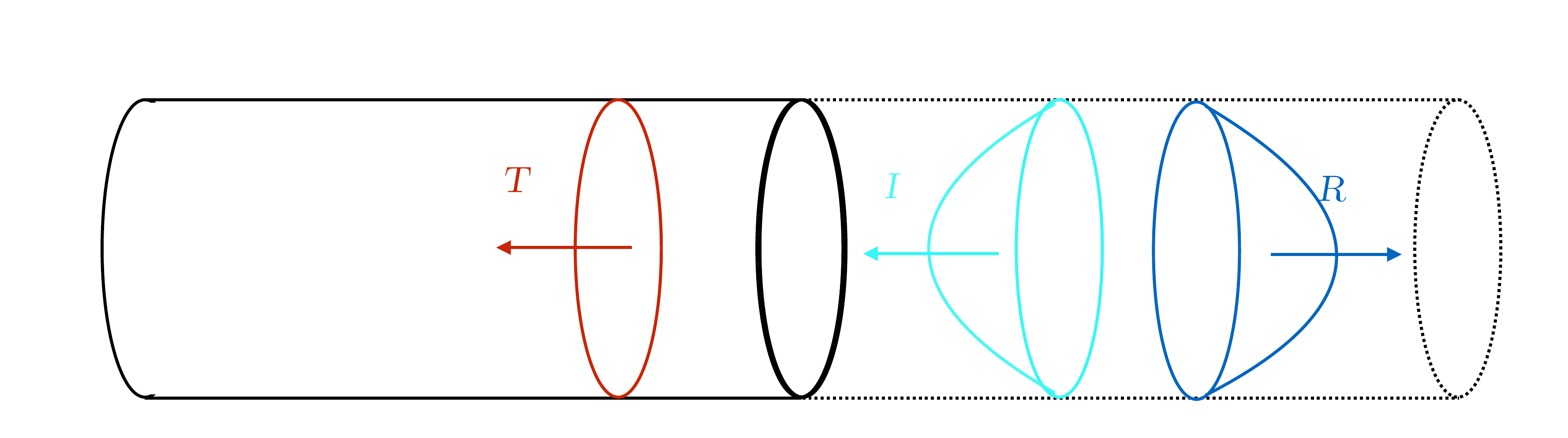}}
\caption{Schematic depiction of simplified jet-nozzle system comprised of connected, solid- and porous-walled cylindrical ducts.}\label{fig:tube}
\end{center}
\end{figure}

Consider the simplified problem depicted in figure \ref{fig:tube}, in which an incident $k^-$ wave of amplitude $I$, impinging on the nozzle plane produces a reflected wave of amplitude $R$, and a transmitted wave of amplitude $T$. Consider mass and momentum conservation in a thin disk (much smaller than a wavelength, meaning that the flow can be considered incompressible) containing the nozzle exit plane, respectively,

\begin{align}
\int u_1' \text{d}A&=\int u_2'\text{d}A, \label{eq:mass}\\
\int p_1' \text{d}A&=\int p_2'\text{d}A, \label{eq:mom}
\end{align}
where the subscripts $1$ and $2$ refer, respectively, to the upstream and downstream faces of the disk.

The incident, reflected and transmitted waves all satisfy the dispersion relation,

\begin{align}
k^{\pm}=\frac{-\omega M\pm\sqrt{\omega^2-4\alpha^2(1-M^2)}}{1-M^2},
\end{align}
where non-dimensionalisation has been performed using the jet diameter and speed of sound. The radial structures of the waves are given by the eigenfunctions $J_0(\alpha r)$ where $\alpha$ is the non-dimensional radial `wavenumber'.

The $k_d^-$ wave behaves, as shown by \cite{TowneetalJFM2017}, like a wave propagating in a porous-walled duct, and its first radial mode has therefore, $\alpha_j=2.4048$. The transmitted wave propagates into the cylindrical, hard-walled nozzle, either as a plane wave, $\alpha_n=0$, or as a wave with higher-order radial structure, characterised by $\alpha_n=3.8,\dots$. 

The cut-on condition for nozzle modes of radial order $\alpha_n$ is,
\begin{align}
St&=\frac{\alpha_n\sqrt{(1-M^2)}}{\pi M}.
\end{align}
For the Mach number range of interest, $0.6 \leq M \leq 0.82$, the corresponding cut-on Strouhal-number range for the first radial pipe mode, $\alpha_n=3.8$, is well above the frequencies of interest. The incident $k^-$ waves will therefore be transmitted as plane waves.

Using the Fourier-transform convention,
\begin{align}
p(x,r,t)=\hat{p}(r)\text{e}^{i(kx-\omega t)},\\
u(x,r,t)=\hat{u}(r)\text{e}^{i(kx-\omega t)},
\end{align}
the pressure fields inside and outside the nozzle that we are interested in are, therefore, respectively,
\begin{align}
\hat{p}_n(x,r)&=T\text{e}^{ik_n^-x},\\
\hat{p}_j(x,r)&=J_0(\alpha_j r)(I\text{e}^{ik_j^- x}+R\text{e}^{ik_j^+ x}),
\end{align}
where the $\text{e}^{-i\omega t}$ has been dropped for convenience, and the momentum balance at the exit plane, $x=0$, reads,
\begin{align}
T  =\frac{2(I+R)}{r^2}
 \int_0^r J_0(\alpha_j r) r\text{d}r\label{eq:mom}
\end{align}

\noindent In terms of the velocity fluctuation, upstream of the exit plane we have,
\begin{align}
\rho_o\Big(\frac{\partial u^-}{\partial t} + M \frac{\partial u^-}{\partial x} \Big)&=-\frac{\partial p^-}{\partial x},\\
\rho_o (\omega-k_n^-M)\hat{u}\text{e}^{i k_n^- x}&=k_n^- T\text{e}^{i k_n^- x},
\end{align}
at the nozzle exit plane, $x=0$,
\begin{align}
\hat{u}^-&=\frac{k_n^- }{\rho_o(\omega-k_n^-M)}T
\end{align} 
and downstream, considering momentum balance separately for the $k^+$ and $k^-$ components of the fluctuations,
\begin{align}
\rho_o\Big(\frac{\partial u^\pm}{\partial t} + M \frac{\partial u^\pm}{\partial x} \Big)&=-\frac{\partial p^\pm}{\partial x},\\
\end{align}
giving, 
\begin{align}
\hat{u}^+&=\frac{k_j^+}{\rho_o(\omega-k_j^+M)}RJ_0(\alpha_jr),\\
\hat{u}^-&=\frac{k_j^-}{\rho_o(\omega-k_j^-M)}IJ_0(\alpha_jr).\\
\end{align}
The mass balance at the nozzle exit plane is
\begin{align}
\int\hat{u}_1^-\text{d}A&=\int (\hat{u}_2^++\hat{u}_2^-) \text{d}A,\\
\frac{k_n^- }{\rho_o(\omega-k_n^-M)}T r^2&=2\Big[\frac{k_j^+}{\rho_o(\omega-k_j^+M)}R+\frac{k_j^-}{\rho_o(\omega-k_j^-M)}I \Big]\int  J_0(\alpha_jr) r\text{d}r,
\end{align}
giving for the amplitude of the transmitted wave,
\begin{align}
T=2\frac{\rho_o(\omega-k_n^-M)}{r^2k_n^-}\Big[\frac{k_j^+}{\rho_o(\omega-k_j^+M)}R+\frac{k_j^-}{\rho_o(\omega-k_j^-M)}I \Big]\int  J_0(\alpha_jr) r\text{d}r,\label{eq:mass}
\end{align}
Combining equations \ref{eq:mom} and \ref{eq:mass} to eliminate $T$,
\begin{align}
{(I+R)}&=\frac{\rho_o(\omega-k_n^-M)}{k_n^-}\Big[\frac{k_j^+}{\rho_o(\omega-k_j^+M)}R+\frac{k_j^-}{\rho_o(\omega-k_j^-M)}I \Big].
\end{align}
From which the reflection coefficient can be obtained,
\begin{align}
\frac{R}{I}&=\Big(\frac{1-\frac{k_j^+M}{\omega}}{1-\frac{k_j^-M}{\omega}} \Big)\Big(\frac{k_n^--k_j^-}{k_j^+-k_n^-}\Big).
\end{align}

%


\bibliographystyle{mystyle}
\bibliography{biblio4punkss}

\end{document}